# A Novel Predictive and Non-Predictive Cooperative Model for Routing in Ad Hoc Networks

S. A. Sahaaya Arul Mary, Ph.D
Anna University, Chennai, Tamil Nadu, India

Jasmine Beulah G,
Anna University, Chennai, Tamil Nadu, India

## ABSTRACT
Ad hoc networks are formed by intermediate nodes which agree to relay traffic. The link between nodes is broken when a node rejects to relay traffic. Various parameters like depreciation in the energy of a node, distance between nodes and mobility of the nodes play a vital role in determining the node's rejection to relay traffic. The objective of this paper is to propose a novel model that identifies the cooperative nodes forming stable routes at the route discovery phase. The weight factor of the different parameters decides the varied type of networks where the proposed model can be applied. Hence, an Artificial Neural Network based non-deterministic generic predictive model is proposed to identify the varied types of networks based on the weight factor. This study has been substantiated by simulation using OMNET++ simulator. We are sure that this paper will give a better solution to identify cooperative nodes thereby improving the performance of the network.

## 1. INTRODUCTION
In the present scenario, fixed base stations or access points for communication are actively being replaced by mobile wireless environment which has led to the need for Mobile Ad-hoc Networks. A node in a mobile network has a direct communication with its neighbors but adapts itself to some intelligent routing strategy to discover paths to other nodes that provide an efficient and stable communication. Researchers have proposed various algorithms for data transmissions in an ad hoc environment which includes Dynamic Source Routing (DSR) [1] and Ad hoc On Demand Distance Vector Routing (AODV) [2]. These algorithms do not choose a route based on the stability of links but focus on shortest path leading to frequent route discoveries. Frequent route discoveries leads to increase in control overhead, consumes bandwidth, and loss in energy of the nodes thereby leading to poor network performance.

Routing and Network management must be done cooperatively by all the nodes. Different mobile nodes with different goals share their resources in order to ensure global connectivity. However, some resources are consumed quickly as the nodes participate in the network functions. An individual node may attempt to benefit from other nodes, but refuse to share its own resources. Such nodes are called non-cooperative nodes. A non-cooperative node may refuse to forward data packets for other nodes in order to conserve its own energy [3] [4].

A node drops packets when the neighboring node moves away from its transmission range and when it becomes a non-cooperative node in terms of various parameters. So, it is important that the neighboring nodes must be in vicinity with each other and cooperative in nature to relay traffic and Mobile Ad hoc networks operating in critical environment require stable routes which depend on the link in that route. The successful operation of message transfer in a military environment will be affected when an intermediate node with high signal strength moves out of transmission range. The situation becomes even worse when the source and destination having agreed upon a communication link suddenly fails to relay traffic. This sudden rejection to relay traffic by the intermediate nodes may be based on some factors such as the energy constraints of the node, the node's location in the network and the application demand [5].

In MANETs, routing misbehavior is due to non-cooperative nodes which severely degrade the overall performance of the network. Some of the nodes may violate the requirement to relay traffic disrupting the network. The nodes must be monitored and appropriate measures must be taken in order to identify the non-cooperative nodes. Ultimately identifying cooperative behavior of the nodes at the route discovery itself will improve the performance of the network.

Consider the following important questions which help to solve the problems encountered in identifying cooperative nodes.

(a)      What are the crucial parameters needed to be considered for detecting cooperative behavior in different types of networks?

(b)      What is the deviation between expected behavior and actual behavior of nodes?

The answer to the above questions will definitely improve the identification process of cooperative nodes and thereby improve the performance of the network. Neural Network is a popular technique to predict reasons/solutions for various types of problems encountered. This paper presents a routing protocol, CO-AODV (Cooperative Ad hoc On Demand Distance Vector) which identifies cooperative nodes constituting stable routes. This paper also uses the ANN (Artificial Neural Network) technique to identify the cooperative nodes to suit varied types of networks based on the weight factor. We use OMNET++ simulations to identify the stable routes based on cooperative behavior.

The paper has been organized in the following manner. Section 2 summarizes the related work on stable routing. In Section 3, we discuss the motivation to write this paper. Section 4 explains the stability features needed for a node to relay traffic and the proposed algorithm to identify stable routes. Section 5 discusses the deterministic solution and a non-deterministic solution to identify cooperative nodes and thus identify stable routes and the type of network and shows the results in a graphical form. Future directions for the paper are given in Section 6. Finally, we conclude the paper in Section 7.

## 2. RELATED WORK

The problem of route stability has been widely addressed in the literature by considering different parameters. Signal Strength has been used as a metric by several routing schemes in the past research years of Ad hoc Networks. Strongly connected and weakly connected links are distinguished in [6] by introducing a stability threshold. A link is strongly connected if it has been active for a certain period of time. In [7] Shrestha and Mans have pointed out that the neighboring data flows affect the energy drain of a node. Toh in [8] selects a path based on minimum total transmission power and sufficient residual battery power. Network connectivity is preserved by choosing a route based on the remaining battery life of nodes along the route in [9].

The Life-time Prediction Routing (LPR) algorithm in [10] estimates battery lifetime based on the residual energy of the node and its past history. Misra and Banerjee [11] proposed an algorithm which selects a path that has a largest packet transmission capacity. The Critical node's transmission capacity is calculated by taking the residual energy of the node divided by the expected energy spent in reliably forwarding a packet. When a lifetime of a link exceeds a specific threshold which depends on the relative speed of mobile hosts, the link is considered stable in associativity-based routing algorithm [12]. Kozma and William [13] have identified misbehaving nodes based n random audits. Punde et al. [14] considered the availability of nodes by dealing on the improvement of the reactive routing procedure. The stability of the nodes depends on the speed of the node and on the observed packet processing ratio of the given node.

Sridhar and Chan [15] investigated on the analysis of route stability Mobile Ad hoc networks by demonstrating that residual link lifetime is affected by parameters such as speed and mobility patterns. Beraldi et al. [16] suggests that the packet forwarding is not driven by a previously computed path but rather the concerned nodes gather a set of routing meta information called hints to discover the path to the destination. Hints are gathered from the nodes located within a small range of neighbors limited by maximum number of hops.

FResher Encounter Search, a route discovery mechanism in MANETs is proposed by Dubois-Ferriere et al. [17] using encounter ages. Two nodes encounter each other when they are directly connected. FRESH performs a succession of small searches instead of a large route request.

Agarwal et al. [18] suggests a routing protocol called Route-Lifetime Assessment Based Routing Protocol (RABR) where the route selection is done using residual route-lifetime prediction on the basis of link affinity. The signal strength is used as a metric to establish a route with long lifetime. Link Affinity is an estimate of the time after which the neighbor will move out of the threshold signal boundary of a mobile node and hence is a measure of link availability. Lin et al. [19] analyzed the link stability between SSA and ABR. The authors state that in a highly dense environment, shortest path routing finds unstable routes. Targn et al. [20] have considered the radio propagation effect on signal strength to achieve link stability. The authors consider that the link stability is equal to the probability of the receiving signal strength exceeding a pre-defined threshold. A number of challenges have to be overcome to realize the practical benefits of ad hoc networking. From the literature we are able to understand that all the crucial parameters like energy, distance and mobility are not considered to identify route stability. Hence, we propose a deterministic cooperative model to identify cooperative nodes which constitute stable routes.

In [22], the authors use fuzzy neural networks for the construction of management routing subsystem in mobile ad hoc networks. The authors in [23] estimate stable paths in MANETs using mobility prediction. There arises a need to identify cooperative nodes to relay traffic towards identifying stable routes in MANETs and the need to identify the crucial parameter in different types of networks such as wireless networks, MANETs, VANETs and sensor networks. Hence this paper is an extension of the work proposed in [21].

## 3. CO-AODV – MOTIVATION

The main purpose of the research is to identify stable routes for sending and receiving packets between the source and the destination. The energy of the nodes is a very critical resource in an ad hoc environment. So, the nodes try to utilize their energy in a selfish way. The nodes which have very less energy are unwilling to relay traffic. Our main target is to implement a Cooperative AODV in order to achieve high performance in a highly ad hoc environment. The measure of stable path between source and destination is possible when every node in the network store information about their cooperative neighbors. Every node periodically sends "HELLO_PACKETS" to their neighbors and gets updated about their Cooperative Status through the existing AODV protocol. Route Discovery is initiated by the source and the ROUTE_REQUEST packet is broadcasted to its neighbors. The receiving node appends its cooperation status information (TRUE/FALSE) and rebroadcasts it until destination is reached. The hop count along with the cooperative status of the path is used by the destination to choose the cooperative stable path. The ROUTE_REPLY packet is sent by the destination along the identified Cooperative Stable path.

## 4. STABILITY FEATURES

The path contains ordered nodes $\{N_0, N_1 ...N_m\}$ in an ordered set of consecutive links $\{L_1, L_2.....L_m\}$. Let us consider the link Li between two consecutive nodes Ni-1 and Ni on the path. Our aim is to model the availability of network elements in identifying a stable route for transmission. Nodes exhibit cooperative behavior to relay traffic based on three important elements:

(a)The energy of the node is greater than a certain threshold willing to relay traffic.

(b)There exists a Potential Communication capability between successive pair of nodes.

(c)The mobility factor of the nodes should be as less as possible.

In this paper, we assume that the energy constraints of a node, mobility factor and potential communication capability between successive pair of nodes determine the node's cooperative behavior in the link.

## 4.1 Energy Constraints of the Node

Energy of a node is a critical resource which includes three parts such as data sensing, data processing and data reception/transmission. Energy spent on data communication is the most critical cost. Hence, the nodes try to conserve their energy to utilize them for transmission of their own data and thus become non-cooperative. In sensor networks, the sensor node is powered by battery which works for several months without recharging. Hence, in stationary sensory applications, it is important to identify routes which contain cooperative nodes in terms of energy. So, in this type of network, high priority will be given to energy and subsequently mobility and distance is considered.

## 4.2 The Potential Communication Capability of the Node

The potential communication capability of a node depends on the Manhattan distance between the nodes. For a node n1 (transmitting node), the nodes which are closer to n1 will receive stronger signals and those which are far apart from n1 will get weaker signals.

Equation (1) gives the calculation of Manhattan distance (MD) between two successive pair of nodes be justified, not ragged.

$$MD(N_{i-1}, N_i) = MOD(X_{i-1} - X_i) + MOD(Y_{i-1} - Y_i) \qquad (1)$$

We assume that a node N2 is a strong neighbor of N1, if the Manhattan distance between N1 and N2 is less than or equal to r given by (2),

$$0 \leq MD(N\_(i-1,) N\_(i)) \leq r \qquad (2)$$

Where MD is Manhattan Distance and r is the radius.
Hence, we can say that the potential communication capability between pair of nodes exists when MD (Ni, Ni+1) <= r.
Let the potential communication capability between successive pair of nodes be R (Ni, Ni+1) be a binary random variable. Hence we have equation (3),

R (Ni − 1, Ni) =
$$\begin{cases} 1, potential\ communication\ Capability\ is\ true \\ 0, otherwise \end{cases} \qquad (3)$$

In wireless networks, distance between the node and the base station or node is important and hence for communication between successive pair of nodes to exists, distance parameter will be given high priority. The energy of the node and mobility are given the next priority for routing.

## 4.3 Mobility of Nodes

The mobility of the nodes is decided based on the database which contains the history of the node movement for the elapsed time ET. The mobility will be computed using (4),

$$Mobility = \frac{1}{T}\sqrt{(X_2 - X_1) - (Y_2 - Y_1)} \qquad (4)$$

The symbols X1, X2, Y1 and Y2 are the co-ordinates of the node at time T1 and T2 where T = MOD (T1- T2) In order to state the nodes as co-operative the mobility should be zero or very negligible as possible, since nodes with higher mobility will collapse the discovered route which will make the path highly unstable.

In MANETs and VANETs, mobility parameter is given high priority. A network with high mobile nodes cannot exist. Hence, weight age for mobility will be given higher preference when compared to energy and distance.

## 4.4 Cooperative behavior of Nodes

In an arbitrary time interval, the nodes Ni-1 and Ni are cooperative if and only if there exists a potential communication capability between successive pair of nodes, and there is a least mobility between nodes and the residual energy of nodes are greater than a certain threshold. The criteria as shown in (5) (6) (7) should be satisfied in order to state the node as co-operative.

$$E < e, \quad \text{E is energy level of node,} \qquad (5)$$
$$MD < R \quad \text{MD Manhattan distance,} \qquad (6)$$
$$\text{R is radius}$$
$$Mobility = 0 \text{ , least mobility among other nodes} \qquad (7)$$

## 4.5 CO-AODV Algorithm

The algorithm to identify cooperative nodes and thereby selected the best stable link is as follows:

**CO-AODV: Route Discovery Procedure**

1: Initialize set of nodes as M.
2: Start with S.
2: For each Neighbor Ni of S do
    If Mobility (Ni) < Least (Neighbors of S)
      If E (Ni) > = E_Thres
        If MD(S, Ni) < R
          {
Update the value in ROUTE_REQUEST Packet.
          Cooperative CO=1;
        }

**At the Destination Node:**

  1: For each Link Li,
Check whether CO=1 in the ROUTE_REQUEST Packet.
2: If CO=0,
      Drop Li
  Else
      Add (Li, CL)    // add to Cooperative Link, CL
The above procedure identifies the cooperative node in the Route Discovery procedure itself. Various links which are stable are identified. The destination chooses the stable link with the lowest hop count and the stable route is identified and sent as ROUTE_REPLY packet back to the source.

## 5. EXPERIMENTAL SECTIONS

To implement the research study the ad hoc nature of the network structure has been considered. This structure takes 10 nodes as a sample set. This has been simulated using OMNET++ simulator and results are obtained.

## 5.1 Deterministic Solution

The simulation parameters found in Table I have been considered for the sample of size 10 nodes.

**Table1. Simulation Parameters**

| Parameters | Values |
|---|---|
| N(Number of Nodes) | 10 |
| Space(area) | 100 X 100 |
| Tr(Transmission range, units) | 20 |
| Threshold Energy(in units) | 500 |

The Fig.1 shows the ad hoc network structure comprised of 10 nodes as the size of sample. The node N0 and N9 have been considered as source and destination respectively. This wireless ad hoc nature of network efficiency based on the distance between the nodes involved in communication process. The link carries the distance between the pair of nodes. The AODV protocol considers only distance metric to decide the path between the source and destination.

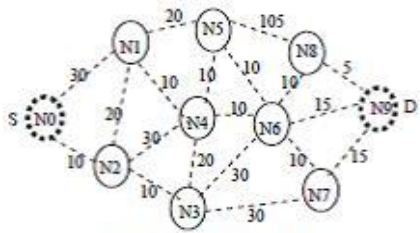

Fig 1 Ad hoc Network Structure

Table II shows the mobility value of each node in the database maintained based on the present and previous co-ordinate of the nodes.

**Table II. Mobility Database**

| Nodes | Position at T1 | Position at T2 | Mobility |
|---|---|---|---|
| N0 | (10, 10) | (10 ,10) | 0 |
| N1 | (20 ,30) | (20 ,31) | 1 |
| N2 | (20 ,10) | (20 ,10) | 0 |
| N3 | (30 ,10) | (34 ,10) | 4 |
| N4 | (30 ,30) | (33 ,30) | 3 |
| N5 | (30 ,40) | (31 ,40) | 1 |
| N6 | (35 ,35) | (38 ,38) | 6 |
| N7 | (40 ,30) | (44 ,30) | 4 |
| N8 | (40 ,40) | (40 ,40) | 0 |
| N9 | (45 ,40) | (46 ,40) | 1 |

The nodes energy and mobility value from the database are shown in the Table III at the time T1.

**Table III. Nodes Energy and Mobility**

| Nodes | Energy | Mobility |
|---|---|---|
| N0 | 1000 | 0 |
| N1 | 500 | 1 |
| N2 | 600 | 0 |
| N3 | 300 | 4 |
| N4 | 700 | 3 |
| N5 | 1600 | 1 |
| N6 | 1500 | 6 |
| N7 | 300 | 4 |
| N8 | 600 | 0 |
| N9 | 400 | 1 |

**AODV Route Discovery:** At the initial stage all nodes are exchanging hello messages. The distance from source to neighbor nodes towards is identified and node which exists at shortest distance from the source will be considered. Similarly, the same steps repeated till the destination node has been reached.

**CO-AODV Route Discovery:** This improved method finds the co-operative node in the path towards the destination from the source. This co-operative node has been decided based on the energy level, mobility and distance factor. If more than one node falls within the same communication range of the node then the node with less mobility and higher energy will be considered. Otherwise the node which has the energy beyond the threshold will be considered for relaying traffic. The route request message carries a Route Stability field which will be marked if the node is co-operative. The Table IV shows that the ROUTE_REQUEST packet towards the destination by the source node. This carries the nodes which are co-operative in providing the communication.

**Table IV. Route_Request Packet**

| Source | Destination | Route Stability |
|---|---|---|
| | | |

The Table V shows the communication path from the source N0 to destination N9. The proposed method finds out the nodes which are highly co-operative to ensure high packet delivery ratio. Since the Co-operative property is directly propositional to packet delivery. This in turn increases the throughput.

The Fig.2 shows the hop counts made by the AODV and CO-AODV.

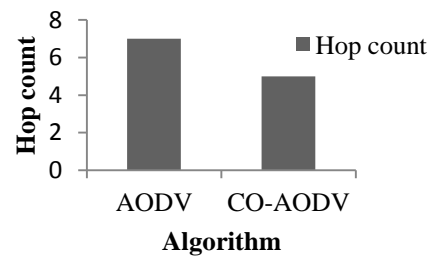

**Fig.2 AODV Vs CO-AODV**

It has been realized that proposed algorithm takes less hop count leads to less propagation time towards reaching destination node.

## 5.2 Non-Deterministic Solution: ANN Based Model

The deterministic solution considers all the parameters in a fixed static way. But this may be varying. Thus, Artificial Neural Network (ANN) based model has been proposed to consider all the possibilities to give generic solution which suit varied types of networks. Fig.3 shows an ANN based solution to identify the co-operative nodes of the protocol to suit varied types of networks. The cooperativeness of the node has been decided based on distance (D), Mobility (M) & Energy (E). But the weight decides the network type where this proposed protocol can be applied.

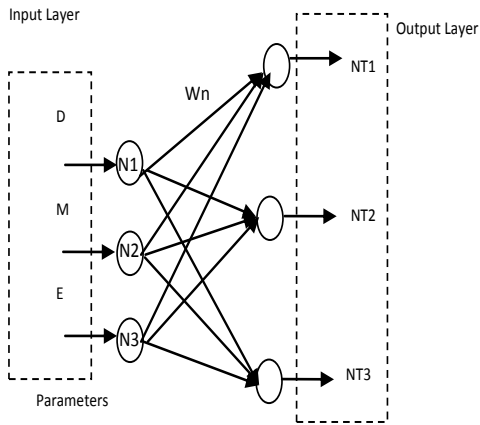

**Fig.3 ANN based Model**

NT1 gives maximum weight to distance.
$$NT1 = D*w1 + M*w2 + E*w3$$
$$\{w1 = 0.5,\ w2 = 0.25,\ w3 = 0.25\}$$
NT2 gives maximum weight to Mobility.
$$NT2 = D*w1 + M*w2 + E*w3$$
$$\{w2 = 0.5,\ w1 = 0.25,\ w3 = 0.25\}$$
NT3 gives maximum weight to Energy.
$$NT3 = D*w1 + M*w2 + E*w3$$
$$\{w3 = 0.5,\ w1 = 0.25,\ w2 = 0.25\}$$

Table VI shows the parameter values and corresponding network type gets selected by the ANN based Model.

**Table VI. Parameter Values**

| D | M | E | Network Type Selected |
|---|---|---|---|
| 0.5 | 0.25 | 0.25 | Distance Sensitive Network (Eg. Wireless Networks) |
| 0.25 | 0.5 | 0.25 | Mobility Sensitive Network (Eg. MANETs,VANETs) |
| 0.25 | 0.25 | 0.5 | Energy Sensitive Network (Eg. Sensor Networks) |

This method also suggests the different type of network where this protocol can be applied by changing the weight metric associated with the parameters.

## 6. FUTURE DIRECTIONS

This study considers very few essential parameters on deciding the co-operative nodes. This can be extended further by incorporating some more vital metrics.

## 7. CONCLUSION

This study has considered the effective data transmission between the pair of nodes involved in communication process. The existing AODV protocol lacks in identifying the proper nodes in order to give good throughput. Thus, this work has proposed CO-AODV as a protocol over the AODV. It has considered the vital metrics to decide the co-operative nodes at the route discovery process, which ultimately helps to reduce the packet loss, reduced hop counts and increases the packet delivery ratio. An Artificial Neural Network (ANN) based predictive model to identify the co-operative nodes of the protocol to suit varied types of networks based on the weight factor is also proposed. This study has been evidently supported by OMNET++ simulator.